# Spin of the M87 black hole

David Garofalo

Department of Physics, Kennesaw State University, USA

Abstract

Spin measurement of the 6.5 billion solar mass black hole in M87 from the Event Horizon Telescope image is the latest in a series that span a wide range in values, but that tend to share the feature of co-rotation between the accretion flow and black hole. The spin paradigm for black holes predicts very high black hole spin which in that framework was produced in its last significant merger. High black hole spin appears to be ruled out in the gap paradigm, however, which predicts early formation with a mass already in excess of 4 billion solar masses. In this picture, the black hole experienced slow evolution as it departed from its original radio quasar phase and over billions of years not only failed to double its mass but also fell short of regaining its original high spin, such that it is now compatible with a co-rotating accreting black hole whose dimensionless spin fits best in the range $0.2 < a < 0.5$.

Introduction

M87 is a low redshift (z~0.004) radio galaxy in the Virgo cluster about 54 million light years away whose central supermassive black hole is estimated at 6.5 billion solar masses by the Event Horizon Telescope (EHT collaboration 2019), a result that follows two decades of measurements (Macchetto et al 1997; Gebhardt & Thomas 2009; Walsh et al 2013). Its dimensionless black hole spin has been estimated at $a \sim 0.98$ (Feng & Wu 2017), $a \sim 0.9$ (Tamburini et al 2019), $a > 0.8$ (Li et al 2009), $a > 0.65$ (Wang et al 2008), $a > 0.2$ (Doeleman et al 2012), $a \sim 0.1$ (Nokhrina et al 2019), $0.1 < a < 0.5$ (Sob'yanin 2018). Despite a variety of model-dependent assumptions and values of spin, co-rotation between the black hole and disk is preferred. We discuss model constraints stemming from the large black hole mass and the low redshift value of M87.

The spin paradigm for black hole accretion and jet formation emerges from the seminal analytic exploration of jet power from black hole spin-energy extraction (Blandford & Znajek 1977; Wilson & Colbert 1995; Moderski, Sikora & Lasota 1998; Sikora et al 2007; Tchekhovskoy et al 2010). Although the paradigm emerged in the thin accretion disk context, it has evolved into a framework that requires thick disks, at least in the inner parts, in order to model the powerful, collimated jets, that are observed in radio galaxies and quasars. The gap paradigm (Garofalo, Evans & Sambruna 2010), on the other hand, opens a window for the most effective jet formation via retrograde accretion or counterrotation between the disk and black hole. Because the most powerful radio quasars in this model emerge from thin disks, thick disks are not required for jet formation. While both models predict co-rotation for the M87 black hole, we show that they predict non-overlapping regions of the prograde spin space. In Section I we introduce the spin paradigm and the gap paradigm and their different predictions for the spin of the FRI radio galaxy M87. In Section II we conclude.

I. Discussion

- The spin paradigm

The spin paradigm for black hole accretion and jet formation amounts to a collection of ideas dating to the 1970's when accretion power and black hole spin energy extraction were placed into a proper theoretical context (Shakura & Sunayaev 1973; Blandford & Znajek 1977). This led to a picture for powerful jets based on the value of the dimensionless black hole spin (Wilson & Colbert 1995). In the following decades and with support from numerical simulations, the high/low black hole spin dichotomy for jetted/non-jetted black holes was enhanced by the introduction of advection dominated accretion flows (Narayan & Yi 1995) associating jets with thick inner disk regions, necessary for collimation and acceleration. Simulations, however, struggled to produce jet efficiencies that can explain observations (Cavagnolo et al 2010; McNamara & Nulsen 2012). This need to find a more efficient jet efficiency led to the introduction of steeper spin dependence at high spin (Tchekhovskoy et al 2010) and flooded magnetospheres (McKinney et al 2012). Even in the latter context where black holes are drowned in magnetic fields by construction, jet efficiency is high enough to match observations only at high spin. The reasons for this may have something to do with the magnetic tower effect (Lynden-Bell 2003; Meier 2012). Retrograde accretion, on the other hand, appears more amenable to the magnetic tower effect (Meier 2012).

While radio quiet AGN are prescribed to inhabit a black hole spin range $0 < a < 0.15$, radio loud AGN instead produce jets whose power is proportional to black hole rotation to the sixth power for thick disks, a much steeper dependence than in the original Blandford-Znajek effect (Tchekhovskoy et al 2010). For the black hole in M87 which has a powerful FRI radio jet extending thousands of parsecs from the center of M87 observed at 14 degrees from our line of sight, these ideas require its black hole to be spinning in co-rotation with its disk at more than 90% of its maximum possible value. General relativistic simulations have also explored disk orientation, showing that the prograde accretion regime produces more efficient jets than the retrograde one (Tchekhovskoy & McKinney 2012). From the perspective of general relativistic simulations, an even higher retrograde spin than prograde spin could match jet power so the retrograde regime is not entirely ruled out for M87. Simulations have also recently made strides towards implementing radiative losses (e.g. Ryan et al 2017) but the impact of this is too early to assess. Despite averaging over extended spatial scales and simplifying assumptions about thermal particle distributions, the results of these simulations tend to be applied prematurely to observations despite the fact that small-scale physics clearly matters (Kunz et al 2016). The possibility that retrograde accretion may be more efficient in producing jets than in prograde ones has been explored analytically (Garofalo 2009; Garofalo 2017). While a much larger range of prograde spin values is compatible with the gap paradigm, the high spin range instead appears to be ruled out as we now show.

- The gap paradigm

The gap paradigm for black hole accretion and jet formation amounts to a collection of ideas that were combined in 2010 including flux trapping (Reynolds et al 2006), retrograde accretion (Garofalo 2009) and eventually grounded in the size of the gap region between the accretion disk and the black hole horizon (Garofalo, Evans & Sambruna 2010). Applying basic phenomenology to the Blandford-Znajek and Blandford-Payne mechanisms (Blandford & Payne 1982), the gap paradigm prescribes FRII jet morphology is formed in retrograde configurations as a result of the large gap region between the inner disk and the black hole horizon and FRI jet morphology for corotating configurations and mostly advection dominated accretion. Because retrograde accretion tends to be unstable, a restrictive parameter space that connects the angular momenta of the black hole and accretion disk (King et al 2005; Sesana et al 2014), and indirectly the masses, has been obtained (Garofalo, Christian & Jones 2019). These conditions make it so that retrograde accreting black holes constitute a minority among all accreting black holes. The reason why the state of accretion does not affect the presence of the jet in retrograde configurations is due to the absence of disk suppression in retrograde states. Instead, the suppression of the jet by the disk occurs in radiatively efficient disks whose inner edge is close to the black hole, hence in the higher prograde spin regime. A fundamental feature of the gap paradigm that is inevitably incorporated into the model is the time evolution which is dictated by the accretion process. Because retrograde configurations spin black holes down, there is a sequence of states that emerge in the paradigm, namely prograde accretion states in the future of retrograde accretion states. This does not mean that prograde accretion states cannot make up initial configurations, simply that if they do constitute initial states, they will also constitute future states. In other words, the cause-and-effect relation is in the sequence retrograde -> prograde and not the other way around.

Whereas the retrograde to prograde evolution is an inevitable consequence of accretion, jet feedback can affect the future accretion state and therefore contribute to determining whether the disk remains thin or evolves into an advection dominated flow. These differences have been applied to explain a host of recent observations (Garofalo et al 2019; Garofalo & Singh 2019; Garofalo, Singh & Zack 2018; Garofalo et al 2016). For our purposes, we focus on the evolution of objects that can model M87, which amounts to an originally retrograde accreting black hole surrounded by a radiatively efficient thin disk that was formed after a merger and that evolved via accretion (Figure 1). This powerful FRII quasar (Figure 1 lowest panel) produced a jet feedback that dramatically affected the entropy of the ISM and eventually altered the mode of accretion on relatively short timescales compared to systems whose jet feedback is weaker. As a result, the accretion disk evolved into a hot, advection dominated state, within only a few million years (Figure 1 second to lowest panel). Of crucial importance here is the state transition of the disk, which as an ADAF accretes at or below 0.01 times the Eddington accretion limit, the theoretical boundary between advection dominated and radiatively efficient disks.

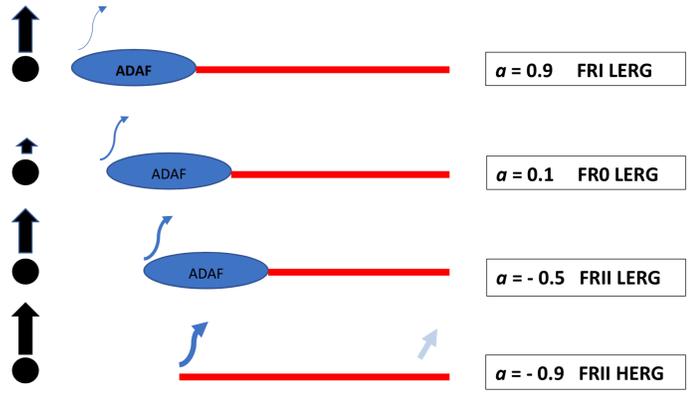

Figure 1: Time evolution of a high spinning black hole surrounded by a counterrotating accretion disk as initial state. The black hole is then spun down toward zero spin and subsequently back up. However, co-rotating disks require much longer timescales to spin black holes up compared to spin down phases. Going from the high spin counterrotating state to a high spin corotating state at the Eddington limit requires upwards of 100 million years. In ADAF states this timescales is enhanced by at least two orders of magnitude (Figure from Garofalo & Singh 2019). LERG refers to low excitation radio galaxy while HERG to high excitation radio galaxy. Spin is also labelled in the right column with negative values representing counterrotation and co-rotation with positive values.

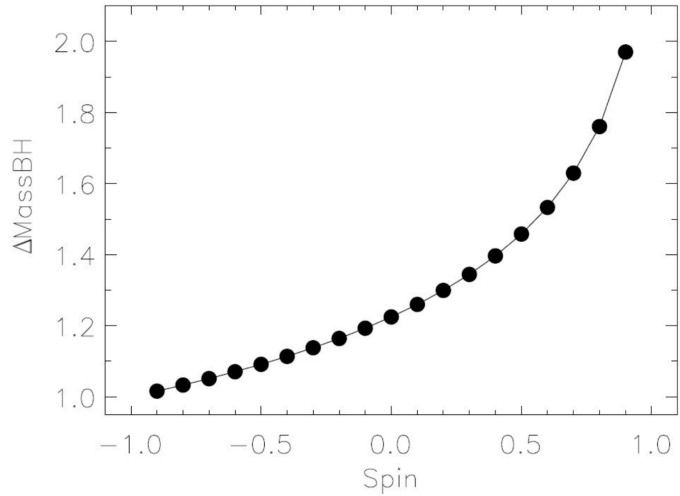

Figure 2: The amount of mass added to the black hole in terms of its original mass as the spin evolves from the high retrograde value (at -0.9) to prograde values. The black hole mass roughly doubles (From Kim et al 2016).

Continued accretion spins the black hole down to zero spin and eventually into co-rotation with the disk. When the spin is prograde but still too low for a powerful jet to form (Figure 1, second panel from top), we have an FR0 radio galaxy (Garofalo & Singh 2019) and continued accretion turns such objects into FRI radio galaxies like M87 as a result of the increase in spin. As mentioned, the crucial thing to note is that 100 million years is the time required to spin the black hole up to high co-rotating values. However, the disk rapidly transitioned to an ADAF (the FRII LERG state in Figure 1) which means continued accretion will require 100 times as long which brings the timescale to 10 billion years. Because we assumed an accretion rate that is at the boundary between ADAF and thin disk, this amounts to a lower limit on the timescale. A more reasonable estimate for that time T would be 10 billion years < T < 100 billion years. Dry mergers if anything would increase the timescale as they would offer the possibility of resetting the clock if the inner accretion ends up retrograde. The hot gas in which the system is embedded would continue to produce an ADAF onto the black hole which would be the equivalent of a system similar to the second panel from bottom in Figure 1. As a result, the idea that the black hole spins rapidly in co-rotation with the disk appears unlikely in this model. However, M87 displays a powerful FRI jet which means its spin is not only prograde but also above about 0.2. Note also from Figure 2 that the amount of mass that needs to be accreted in the prograde regime to increase the spin by 10 % increases as the spin increases. If the mass increased by 1.5 its original value, Figure 2 shows that its spin becomes about 0.5. This means its original mass is $4.33 \times 10^9$ solar masses. Let us estimate the accretion rate needed to spin the black hole up to a dimensionless spin of about 0.5 as an ADAF and in turn determine the time for that process. The Eddington accretion rate in terms of solar masses per year (from Di Matteo et al 2003) is

$$dM/dt_{Edd} = 2.2 \times 10^{-8} \, M_{BH} \, yr^{-1}.$$

If the black hole entered the ADAF phase with $4.33 \times 10^9$ solar masses its average Eddington accretion rate to build its black hole to $6.5 \times 10^9$ solar masses is 119.3 solar masses per year. But as an ADAF it must accrete below 1.193 solar masses per year. If we assume slightly below that value at 0.002 times the Eddington accretion rate, the required buildup time is

$$T = M_{acc}/(dm/dt)_{avg}$$

where $M_{acc} = (6.5 - 4.33) \times 10^9$ solar masses and $(dm/dt)_{avg} = 2 \times 10^{-3} \times 119.13$ solar masses/year, which gives 9.1 billion years.

If, on the other hand, we assume the black hole in M87 currently has high prograde spin and the mass therefore doubled, it went from 3.25 billion solar masses to 6.5 billion solar masses and therefore accreted 3.25 billion solar masses. The average Eddington accretion rate in this case is 107.25 solar masses per year. At 0.002 times the Eddington accretion rate, the required time to reach the 6.5 billion solar mass threshold is 15.1 billion years. Recent estimates of the accretion rate onto the black hole in M87 are orders of magnitude smaller at about $9 \times 10^{-4}$ solar masses per year (Kuo et al 2014). This is $\sim 10^{-5}$ the Eddington rate. GRMHD simulations for the Event Horizon Telescope Collaboration estimate similar accretion rates (i.e. $\sim 2 \times 10^{-5} \, dM/dt_{Edd}$ : Akyiama et al 2019). For the model to work, the average accretion rate must be above $10^{-3} \, (dM/dt)_{Edd}$. From the model perspective, we can identify a decrease in the accretion rate over time, making the model compatible with these estimates at late times. Figure 1, in fact, shows that M87 originates in the model as a powerful FRII quasar which means its accretion rate was close to Eddington and that it evolved from a HERG to a LERG in a few million years. In other words, it could have transitioned early to accretion rates on the order of $10^{-2}(dM/dt)_{Edd}$ and lingered in the range $10^{-2}(dM/dt)_{Edd} > dM/dt > 10^{-3}(dM/dt)_{Edd}$ for billions of years before dropping further to accretion rates below $10^{-4}(dM/dt)_{Edd}$. Our constraint is on the average accretion rate, not the instantaneous one. From the perspective of our evolutionary picture for the accretion rate, our choice of an average accretion rate of $2 \times 10^{-3} \, (dM/dt)_{Edd}$ appears reasonable, which in turn tells us that the idea of the black hole in M87 having a spin that is outside the range $0.2 < a < 0.5$ produces tension with the model.

If the black hole in M87 increased by about 1.5 times the original black hole mass through ADAF accretion, the original M87 black hole about 9 billion years ago was already near 4.3 billion solar masses in this model, at the upper end of the black hole mass scale, which means that in the model we have anti-hierarchical growth for this system. Both the timescale for evolution into an FRI ADAF and the original black hole mass of M87 suggest that it was one of the massive black holes that formed early in the universe. Although a very high retrograde spin for M87 has not been ruled out by observations, as mentioned above, high retrograde/prograde spin in the gap paradigm is fundamentally connected to jet morphology. And since M87 appears to have a classic FRI jet, it cannot be a retrograde accreting black hole in that model.

Although this paper compares two different paradigms, the hope is that analytic and numerical work will eventually converge and signs of this possibility are coming to light. Not only is much of the physics of black hole accretion still absent in GRMHD, even with the currently implemented physics, we are still learning how to simulate accretion around black holes. The detailed dependence of jet power on the magnetic field strength threading the horizon, black hole mass, and black hole spin, is work in progress. While the simulations are now fully 3D, including radiation remains challenging and the efficiency of black hole accretion therefore uncertain (Morales Teixeira et al 2018). Even basic processes for the transport of angular momentum may not operate as thought previously. In fact, magnetic instabilities may contribute to the accretion process such as in MADs where the magneto-rotational instability is marginally suppressed (Marshall et al 2018). Recently, for example, it was discovered that GRMHD simulations were not sufficiently resolved to appropriately explore dynamo behavior in the accretion flow (Liska et al 2019: 2018arXiv180904608L). This is not a second order effect as it is directly related to the buildup of magnetic flux on the black hole which in turn determines whether a jet is formed.

With techniques that allow radiative effects to be incorporated in increasingly realistic ways, recent GRMHD simulations show that a disk wind may help to collimate the jet. As a result, a new picture is beginning to emerge from GRMHD which to some degree is parting ways with the basic picture in which thick disks are required to accelerate and collimate the jet. Instead, their purpose now seems to be anchored to bringing magnetic flux to the black hole. This need to drag sufficient magnetic flux to the black hole has long been recognized as fundamental, with the caveat that the gap paradigm, by contrast, accomplishes this via the zero flux boundary condition (i.e. the Reynolds condition) in the gap region. The prediction is that as more physics is included and implementation techniques improve, GRMHD simulations will show the jet efficiency to increases in the counterrotating accretion regime. This higher jet efficiency in retrograde systems is crucial in producing the time dependent evolution that has made the gap paradigm fit so well with observations.

II. Conclusions

Whereas both spin paradigm as well as gap paradigm prescribe the black hole in M87 to be in co-rotation with its accretion disk, we have shown that they span mutually exclusive regions of the spin space, the former involving a very narrow high spin range while the latter a larger but lower spin range. No further constraints emerge in the spin paradigm. In the gap paradigm, on the other hand, super Eddington accretion is not a feature of the model which places additional constraints on the evolution timescale. Because the black hole accretes most of its lifetime as an ADAF, billions of years are required to spin it up into a region where the model allows for the formation of a powerful FRI jet. These constraints strongly favor a very high mass for the original black hole that formed in the last significant merger that produced the M87 galaxy. Spin constraints have now emerged in the gap paradigm for all families of AGN but whereas for radio quiet quasars/AGN and radio loud quasars/AGN the spin is high but associated with co-rotating and counterrotating disks, respectively, the powerful FRI radio galaxies appear to fit as co-rotating ADAF disks with intermediate spin values. In other words, the less massive black holes in radio quasars have the highest spins but in retrograde configurations whereas the most massive black holes in radio galaxies tend to have intermediate prograde accreting black holes. Although the physical reasons for these conclusions are different, they also emerge in the simulations of Bustamante & Springel (2019). Current spin estimates in active galactic nuclei include large uncertainties (e.g. Reynolds 2019) but will provide the needed constraints in the foreseeable future. We conclude that the best hope to rule out both of the theoretical ideas discussed here, or at least produce tension with them, is for the spin value of the M87 black hole to be measured in the $0.5 < a < 0.9$ range.

Acknowledgments

DG thanks three referees for detailed criticism that improved the impact of this work.